\documentclass[12pt]{iopart}
\usepackage{graphicx}
\begin{document}

 \title{The origin of diffuse scattering in crystalline carbon tetraiodide}
 \author{L Temleitner and L Pusztai}
 \address{Institute for Solid State Physics and Optics, Wigner Research Centre for Physics, Hungarian Academy of Sciences, P.O. Box 49, H-1525 Budapest, Hungary}
 \ead{temleitner.laszlo@wigner.mta.hu}
 
 \begin{abstract}
 
Total scattering neutron powder diffraction measurements were performed on the tetragonal phase (a=6.4202(5) {\AA}, c=9.5762(12) {\AA}) of $CI_4$. The experiments were followed by Reverse Monte Carlo (for POWder diffraction (RMCPOW)) modeling. Detailed analyses of the resulting particle configurations revealed that the observed diffuse scattering originates from the libration of the molecules. By examining the partial radial distribution functions a distinct carbon-iodine peak at 4.5 {\AA} is found, which appears as a consequence of corner-to-face mutual alignment of two molecules. The occurrence of edge-to-edge alignments is also significant within the first carbon-carbon coordination shell.

 \end{abstract}
 
 \pacs{61.05.fm, 61.43.Bn}
 
  \submitto{\JPCM}
  
 \section{Introduction}

Among the crystalline $CX_4$ halides that do not contain iodine ($CF_4$, $CCl_4$, $CBr_4$), monoclinic ($CF_4$ \cite{cf4a_1969, cf4a_1972, cf4a_deb, cf4b_1993, cf4b_2008}, $CCl_4$\cite{ccl4r_1966} and $CBr_4$\cite{cbr4b_1977}), face-centered cubic ($CCl_4$\cite{ccl4a_1981} and $CBr_4$\cite{cbr4a_1977}) and rhombohedral phases ($CF_4$\cite{cf4b_1993}, $CCl_4$\cite{ccl4r_1966} and $CBr_4$\cite{cbr4r_2008}) are common. Over a long period of time, the structure of carbon tetraiodide has been debated; initially it was classified as cubic (see the reference of \cite{pohl}) or monoclinic\cite{finbak1937}, analogous with the above carbon-tetrahalides. The crystal structure could eventually be solved in 1981, when it was classified as tetragonal ($I\bar{4}2m$), on the basis of single crystal x-ray diffraction data \cite{pohl}. The sample in that work was synthesized by the reaction of ethyl iodide and carbon tetrachloride, using the synthesis of McArthur et al. \cite{synthesis}.

Considering existing knowledge on $XI_4$ type crystals\cite{gei4,gei4_new,sni4a_1923,sni4a_1955,sni4b_1997,xi4_common} that possess a cubic symmetry, this unusual structure has been explained by the close packing of $I$ atoms and by an intermolecular $I-I$ bond\cite{pohl}. The observed large value of the ionization potential, found by ultraviolet spectroscopy, was explained also by the presence of this bond\cite{ups_1987}. Recently, this system was exploited as a model for packing tetrahedra\cite{xi4_common}.

Another interesting feature of this class of materials is the possible presence of plastic crystalline phases, appearing as the rhombohedral phase of carbon tetrafluoride\cite{cf4b_1993}, and the face centered cubic phases of carbon tetrachloride \cite{ccl4r_1966, ccl4plastic} and carbon tetrabromide\cite{cbr4a_1977}. The common feature of these phases is that while the centres of the molecules maintain translational (i.e., crystalline) symmetry, the molecules can rotate (for $CF_4$, see \cite{aston_1960}). This rotation is not free, but hindered by neighbouring atoms. As a result, short-range order exists between ligands belonging to neighbouring molecules. This manifests as diffuse scattering that becomes structured in reciprocal space, due to its relation to the average structure (e.g., for $CBr_4$, see \cite{cbr4_struct,cbr4_transverse,tlcbr4}).

Detailed analyses were presented some years ago for some of the crystalline CX$_4$ compounds, applying Reverse Monte Carlo modeling \cite{pardoccl4,tlcbr4} and molecular dynamics simulation \cite{rey_plastic}. These works showed that correlations between the 'ligands' (Cl atoms in $CCl_4$ and Br atoms in $CBr_4$) have a close relationship with those found for the corresponding liquid phase.

During our systematic studies of crystalline phases of carbon-tetrahalides, diffuse scattering in carbon tetraiodide has been found; this initiated the present study. Since the analysis of the Bragg-pattern has already been done by Pohl\cite{pohl}, the present investigation focuses on pair correlations between atoms and on orientational correlations; concerning these issues, standard crystallographic analysis does not provide complete information on the structural details of the system.
 
\section{Experiment}

The sample has been provided by Aldrich (97\% purity) and was further powdered before the experiment so that random orientation of the crystallites could be guaranteed.

Room temperature carbon tetraiodide has been measured by neutron powder diffraction on the former SLAD diffractometer\cite{slad} (Studsvik NFL, Sweeden). Using the standard wavelength of the instrument (1.119~\AA), the diffraction pattern of the sample could be recorded in the 0.29-10.55~\AA$^{-1}$ range. Scattered intensities from the 8~mm diameter vanadium sample container, background and a solid vanadium rod have also been determined. Standard data normalization and correction procedures\cite{howe_norm} were performed by the \textsc{CORRECT}\cite{correct} computer program; absorption, multiple and inelastic scattering contributions have been taken into account during the process. 

As it turned out later, the sealing of the sample holder has leaked during the measurement, and a small amount of sample has been evaporated from the can. In the pattern several additional Bragg-peaks appeared, which are supposedly originate from the orthorhombic phase of $I_2$ \cite{jod,jod_low} (spontaneous brake-up of $CI_4$ molecules, due to the tight packing of large iodine atoms, is common). The most intense such 'alien' Bragg peak can be observed at 1.7\AA$^{-1}$.

The lattice parameters have been re-determined by the 'Fox' software\cite{fox}: they were found to be a=6.4202(5) {\AA}, c=9.5762(12) {\AA}. The amount of the $I_2$ impurity has been estimated by a joint fit of both compounds using the established structures\cite{jod,jod_low}: the amount was found to be about 2.8\%. This corresponds well with the guaranteed purity of the commercially available compound. This indicates that the scattering from impurities originates from the sample, and not from the accidentally deposited, previously evaporated vapour from the container wall.
 
\section{Structural modeling}

Computer modeling has been performed by using the Reverse Monte Carlo for POWder diffraction (RMCPOW) software\cite{rmcpow}. RMCPOW is capable of simulating both Bragg- and diffuse scattering in Q-space. The molecule has the shape of an ideal tetrahedron, with 2.157 {\AA} bond length between carbon and iodine atoms\cite{electrondiff_1941,hargittai_2001}. The molecular structure was maintained during the simulation by 'fixed neighbours constraints' (fnc)\cite{rmc++}. Carbon-iodine and iodine-iodine distances within the molecules were constrained to the distance regions between 2.1 to 2.2 {\AA} and between 3.42 to 3.62 {\AA} , respectively. These constraints, apart from keeping molecules together, allowed them to rotate by series of single atomic moves. The intermolecular closest approach distances were set at 5.7 {\AA}, 4.0 {\AA} and 3.3 {\AA} for carbon-carbon, carbon-iodine and iodine-iodine pairs, respectively.

Initially, a supercell of $10^3$ times of the Bravais-cell has been constructed, which corresponds to 2000 molecules. Atomic positions in the unitcell were set as published in \cite{pohl}, using the previously determined lattice parameters. At the end of the simulation, a goodness-of-fit $R_{wp}=9.54$\% has been reached and 20 independent configurations have been collected for further analyses. (Between two 'independent particle configurations', each atom in the simulation box has moved successfully at least once.)
 
\section{Results and discussion}

Diffuse scattering from a powdered crystalline sample may result from effects other than positional disorder in the bulk crystalline single phase. Here it is assumed that positional disorder {\it is} the only reason for the observed diffuse scattering; before discussing results from our calculations it is therefore necessary to consider some of the other possibilities. 

First, one might argue that the origin is the molecular iodine (I$_2$) that is present in the sample as impurity; this is not really probable, due to its small contribution and due to the lack of an intense diffuse scattering in the pure phase measurements \cite{jod,jod_low}. Another hypothesis is that diffuse scattering might be related to the (small) size of the grains. However, no broadening of the Bragg-peaks has been observed here that could be related to small grains. As yet another possible reason, the presence of an amorphous phase may be assumed. Amorphisation has been reported for $CCl_4$ (see, for example, \cite{ccl4_amorphous}), which took place during vapor deposition to a cooled substrate. For $SnI_4$ the presence of amorphous phase is reported at pressures over 25 GPa \cite{sni4_am}. In contrast to these (somewhat extreme) cases, the present study was performed at ambient conditions (and on commercial samples) and no trace of any additional amorphous phase could be observed. We therefore can conclude that the observed diffuse scattering is related to bulk phase structural properties of crystalline $CI_4$.

\begin{figure}[p]
\begin{center}
\rotatebox{0}{\resizebox{1\textwidth}{!}{\includegraphics{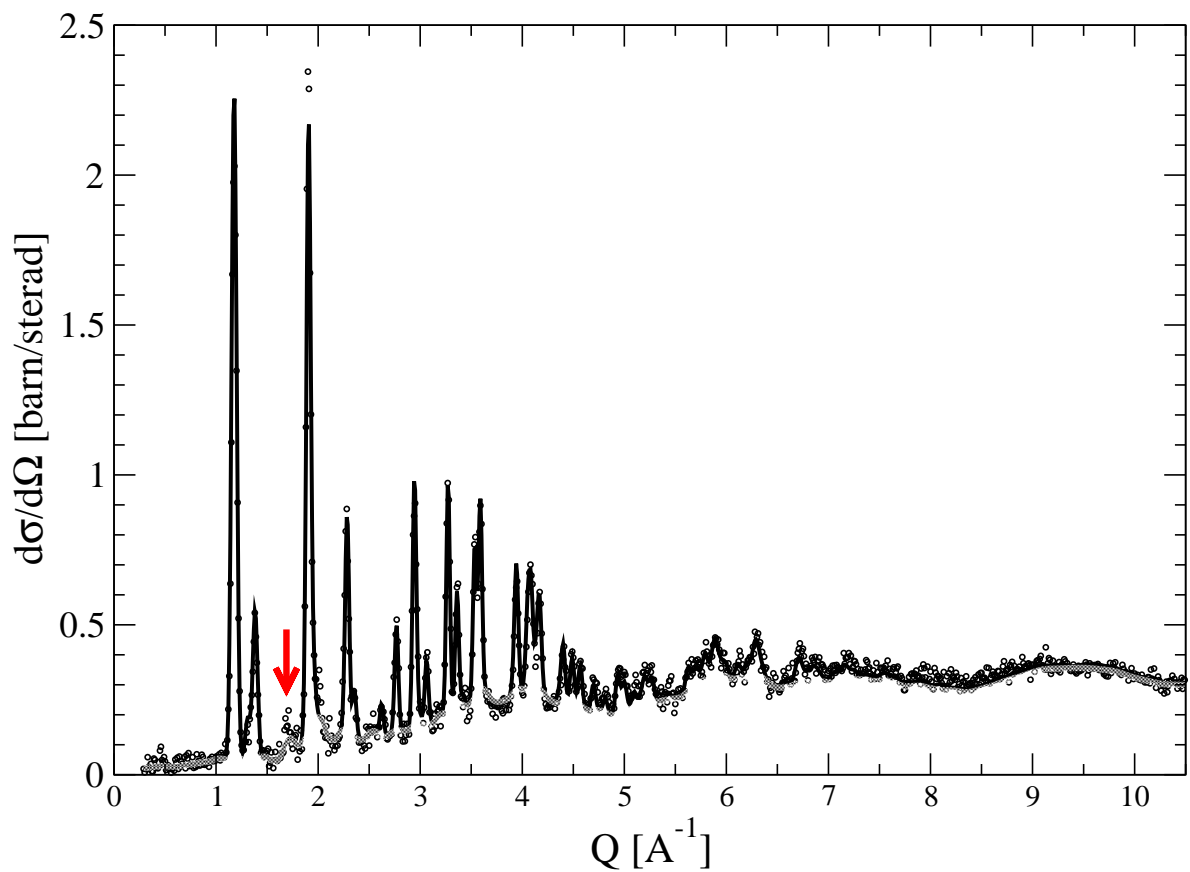}}}
\caption{\label{fig:expsq}
Experimental and simulated powder diffraction patterns of tetragonal $CI_4$ at 298~K. Circles: measured differential cross-section; solid line: RMC calculated total scattering intensities; light red line: RMC calculated diffuse scattering intensities. 
The red arrow marks the most intense parasitic Bragg-peak, supposedly from iodine impurities.
}
\end{center}
\end{figure}

At the start of the simulation, we tried to remove the Bragg-scattering part of crystalline $I_2$ traces from the full pattern, by extrapolating the components of the Debye-Waller factor from the low-temperature measured data\cite{jod_low}. This resulted in a slightly better agreement with experimental data; still, some of the  discrepancies remained, most probably due to the not-well-known magnitude of Debye-Waller factors. This is why we decided not to bias results with this complication and started the simulations without subtracting this kind of background. At the final stage of the RMCPOW calculation, an agreement shown in \fref{fig:expsq} was reached, which (with the R-factor value less than 10 \%) may justifiably be called satisfactory. A possible reason for the remaining differences between measured and calculated data is the presence of Bragg-peaks of the small amount of $I_2$; they are most apparent between 1.5 and 2.5~\AA$^{-1}$ where they caused small modulations of the diffuse scattering part of the simulated pattern.

The coherent neutron scattering lengths of carbon and iodine are 6.646  and 5.28 fm\cite{scattering_lengths}, respectively. Thus the contributions of partial structure factors to the measured total scattering structure factor are 5.7\%, 36.4\%, 57.9\% for carbon-carbon, carbon-iodine and iodine-iodine pairs, respectively. C-I and I-I correlations may therefore be determined well, whereas the position of C atoms (and hence, C-C correlations) is influenced mostly by the applied fixed neighbour constraints. Since none of the 2a and 8i Wyckoff sites restrict the reflection conditions (in comparison with the general 16j Wyckoff site), the above contributions of the partials are valid for all Bragg-reflections, too.

\begin{figure}[p]
\begin{center}
\rotatebox{0}{\resizebox{0.7\textwidth}{!}{\includegraphics{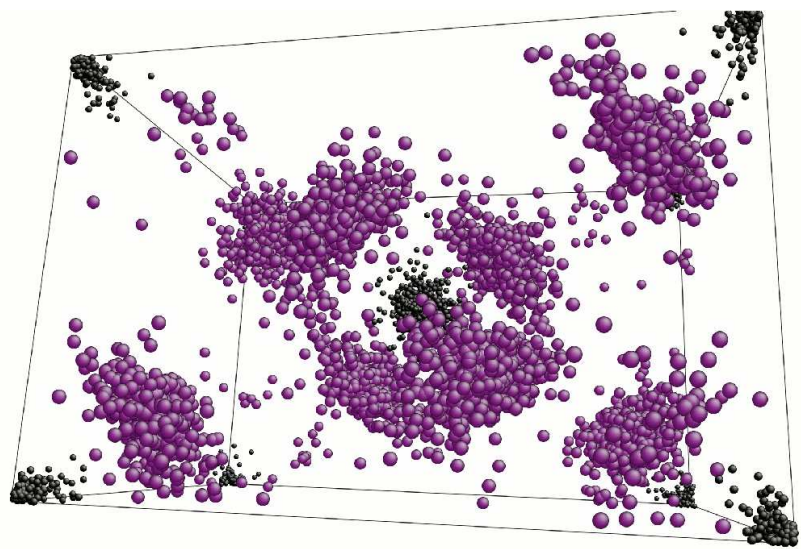}}}
\caption{\label{fig:reduced}
Condensed view of the simulated configuration ($10^3$ unit cells), as projected into one Bravais lattice. Balls represent individual atomic positions. Black: carbon; magenta: iodine. The considerable spread of the iodine positions can be assigned as the major reason behind the observed diffuse scattering intensity. 
}
\end{center}
\end{figure}

\Fref{fig:reduced} shows a condensed view of the supercell as reduced to one Bravais-cell. This picture represents not just the average structure but also the effect of thermal displacements, which latter manifests as the difference between instantaneous and average positions. Intuition based on inspecting the spreads of instantaneous atomic positions around their equilibrium values suggests that the major contribution to the measured diffuse scattering must come from the sizeable displacements of I atoms or, in other words, the libration of the molecules.

\subsection{Partial radial distribution functions}\label{prdf}

\begin{figure}[p]
\begin{center}
\rotatebox{0}{\resizebox{1\textwidth}{!}{\includegraphics{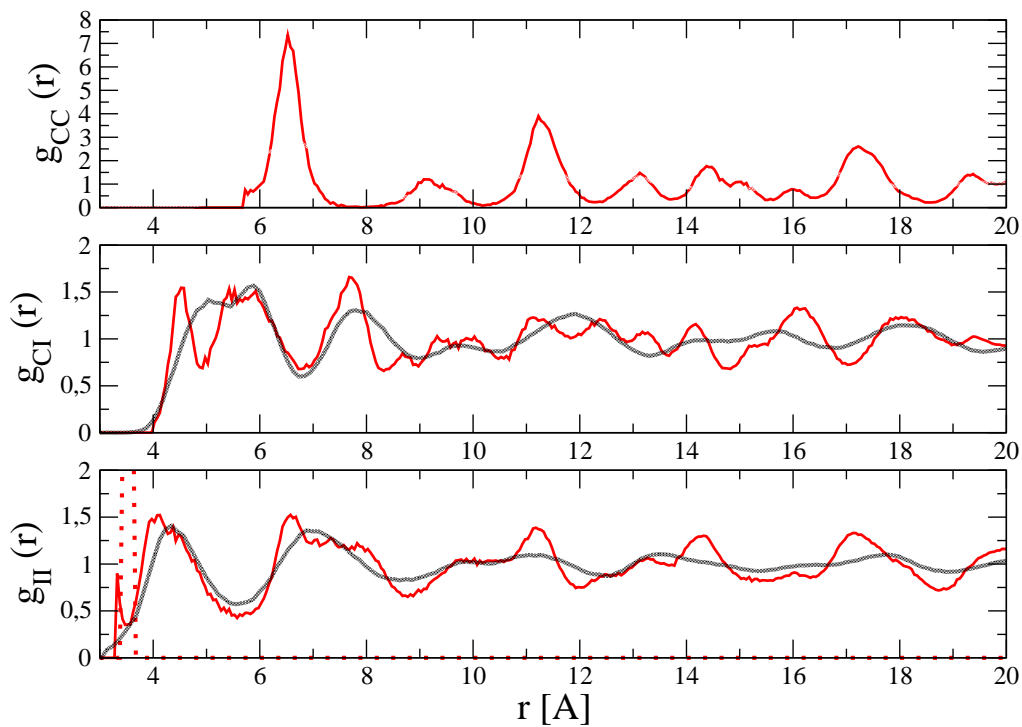}}}
\caption{\label{fig:gr}
Partial radial distribution functions of crystalline $CI_4$ calculated from the simulated configuration. Upper panel: carbon-carbon; middle panel: carbon-iodine; lower panel: iodine-iodine correlations. Red solid line: $CI_4$ RMCPOW intermolecular part; red dashed line: $CI_4$ RMCPOW intramolecular part; light blue solid line: the corresponding RMC simulated\cite{tlcbr4} $CBr_4$ phase II partials (the 'x' axes for these curves have been re-scaled, via multiplying by 1.06 (upper and middle panels) and 1.1 (lower panel).)
}
\end{center}
\end{figure}

Partial radial distribution functions (prdf), as calculated from the final configuration, are shown in \fref{fig:gr}. Among the carbon tetrahalides, the closest analogue of the ordered phase of carbon tetraiodide is the ordered phase of carbon tetrabromide; for the sake of comparison, rescaled prdf-s\cite{tlcbr4} of $CBr_4$ in phase II (monoclinic) are also shown. Since the two structures do not belong to the same spacegroup, exact rescaling (connected to the different sizes of Br and I atoms) is not possible. The scaling factor along the 'x' axis was chosen to be 1.06 for carbon-carbon and for carbon-iodine pairs, taking into account approximate positions of maxima of the carbon-carbon prdf. For iodine-iodine correlations a factor of 1.1 was chosen, taking into account the minimum positions between the first and second maxima.

Comparing the two phases, the carbon-carbon prdf of carbon tetraiodide looks much more ordered; the main origin of this difference is the different crystalline structure (and perhaps the fact that the $CBr_4$ data were measured only about 20 K below the phase transition temperature is also an issue in this respect). The C-C maxima in $CI_4$ can be found approximately at 6.5 {\AA}, 9.1 {\AA}, 11.2 {\AA}, 13.1 {\AA} (and larger distances, too); these values nearly correspond to the shortest lattice side (6.42 {\AA}), the half of the body diagonal (6.59{\AA}), the '$xy$' diagonal (9.08 {\AA}), the second neighbour half body diagonal (11.22 {\AA}), and the full body diagonal (13.15 {\AA}) values, respectively. 

It may be appropriate to discuss here briefly the sharp (unphysical) 'cut-offs' of the $C-C$ and $I-I$ partials at the shortest distances. Usually, this feature may have two sources: (1) the higher than real intermolecular cut-off and (2) non-physical short period oscillations in $Q$-space. Taking into account that the iodine-iodine cut-off distance is lower than the intramolecular low limit, the observed phenomena must be result of that the program tried to fit the Bragg-peaks of $I_2$.

Comparing the centre-ligand and ligand-ligand prdf-s with the corresponding prdf-s of $CBr_4$ phase II, $CI_4$ again appears to be a more ordered system. Concerning the short-range behaviour, I-I maxima are at $4.05$, $4.4$ and $6.6$ {\AA}, where the first and the last ones are distinct. Although the peak at $4.05${\AA} appears, that should possess a crucial role in explaining the formation of the tetragonal lattice \cite{pohl, ups_1987}, it is not so well-defined in terms of pair-correlations. In contrast, the carbon-iodine peak at $4.5$ {\AA} looks much more characteristic. For assigning the observed peaks to local orientations, see \sref{angorcorr}.

\subsection{Angular and orientational correlations}\label{angorcorr}

\begin{figure}[p]
\begin{center}
\rotatebox{0}{\resizebox{0.7\textwidth}{!}{\includegraphics{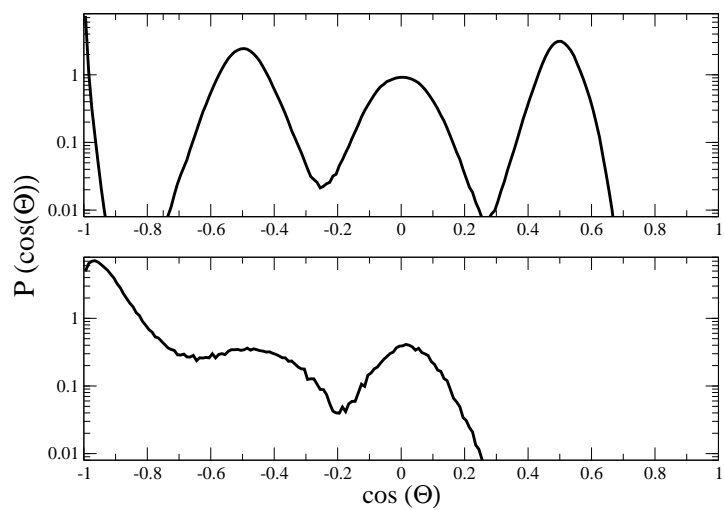}}}
\caption{\label{fig:ang}
Cosine distributions of some intermolecular angles. Upper panel: nearest neighbour C-C-C angle. Lower panel: C-I-C angle distribution. (Note the logarithmic 'y' axes.)
}
\end{center}
\end{figure}

Angular distributions (\fref{fig:ang}) C-C-C and C-I-C were investigated in the following way: atomic triples contributed to the statistics if the distance between carbon atoms were less than 7.5 {\AA} and less than 4.85 {\AA} between carbons and iodines. These values are upper limits of the first coordination shells of the corresponding prdf-s (see \sref{prdf}). For C-C-C triplets maxima at 60, 90, 120 and 180 degrees appear, which correspond well to the local symmetry of the crystal. However, the spread around the regular angles is not the same: the narrowest maximum is the 180 degree one, which shows chain-like ordering of the 'a' sides of the unitcell and through the body diagonal. Less ordered is the body diagonal-'a' side combination (at 60 and 120 degrees) and the broadest one is between the 'a' sides.

For C-I-C correlations, a maximum is found around 165 degrees (-0.965 in cosines) and two less intense ones at 120 and 90 degrees. The first one is at a larger angle than the crystallographic result\cite{pohl} (153.6 degrees (cosine: -0.896)), showing that iodine atoms prefer to be close to, but not exactly on, the carbon-carbon connecting line. Since the intensities of the remaining two maxima are much smaller (note the logarithmic scale in the figure!), supposedly they are spurious intensities arising from C-I correlations that can be found at the high end of the distance range. This kind of correlation is then responsible for the observed peak of the C-I prdf at 4.5 {\AA}. It is instructive to note that if we calculate the average angle from the known maxima of the partial rdf-s (C-C: 6.5 {\AA}, C-I: 4.5 {\AA} and the bondlength of 2.157 {\AA}), then we obtain almost the same result as the one derived from crystallography. That is, the average value is actually somewhat misleading here; clearly, the distribution is the proper characteristics of the structure.

For describing orientational correlations between pairs of tetrahedral molecules as a geometrical unit, the classification scheme of Rey\cite{rey} has been used. This is based on the number of atoms of each of the two molecules found between two parallel planes, which planes contain the centres of the corresponding molecules and are perpendicular to the line connecting the two centres. Based on this consideration, six classes exist: 1:1 (corner-corner), 1:2 (corner-edge), 2:2 (edge-edge), 1:3 (corner-face), 2:3 (edge-face), 3:3 (face-face). The asymptotic limits of the probabilities of 1:1 and 3:3 are the same; this is also true for the 1:2 and 2:3 arrangements. These equivalences come from the fact that it should be 4 atoms on average between the two planes. In the case of tetrahedral liquids this resulted in an alternating behaviour between the 1:1 and 3:3, and between the 1:2 and 2:3 orientation probabilities (c.f. Refs. \cite{rey, rey_common, szilviccl4, ccl3br_caballero, cbr2cl2}). Similar behaviour was observed in the plastic crystalline phases of the corresponding materials\cite{tlcbr4,ccl3br_caballero}.

\begin{figure}[p]
\begin{center}
\rotatebox{0}{\resizebox{0.7\textwidth}{!}{\includegraphics{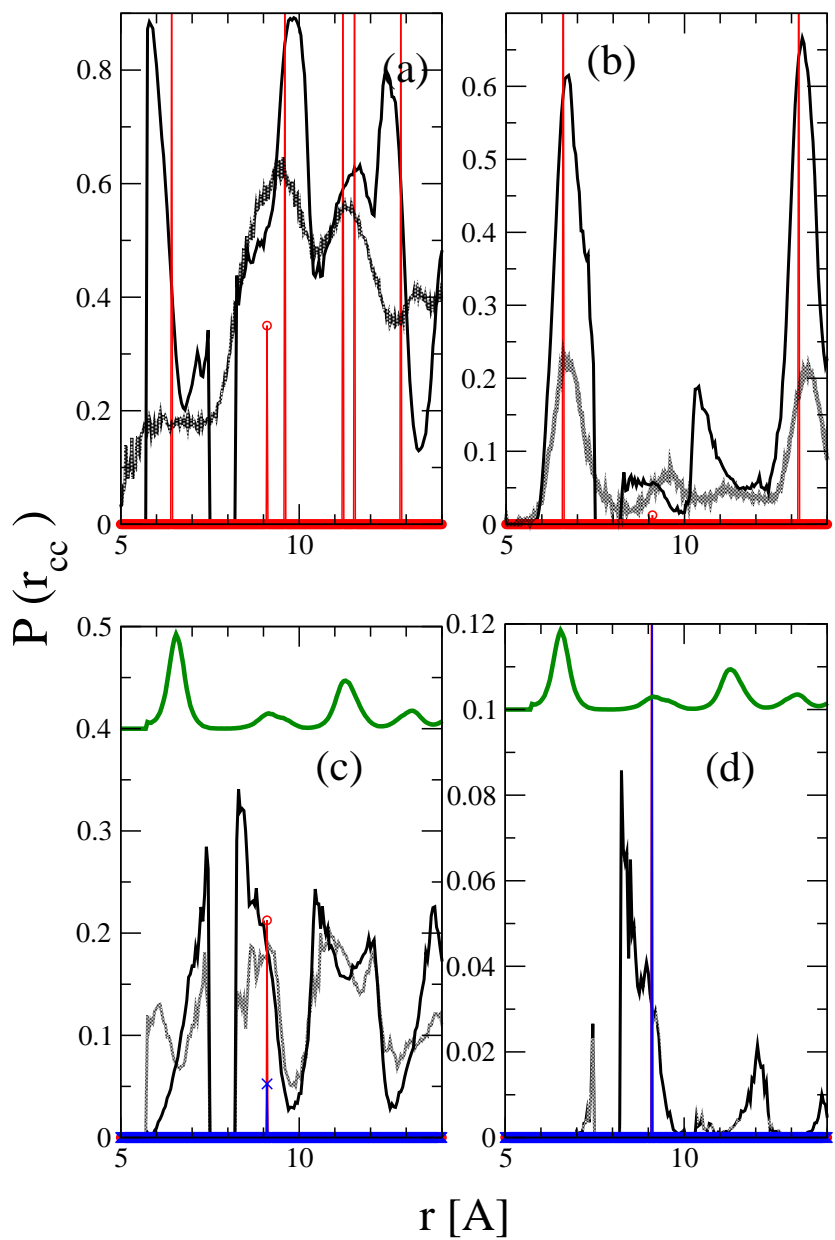}}}
\caption{\label{fig:rey}
Rey's constructions\cite{rey} of orientational correlation probabilities as functions of the centre-centre distance. Upper panels [(a) and (b)], black solid lines: 2:2 [(a)] and 1:3 [(b)] arrangements; red solid lines: the corresponding probabilities as calculated from the average structure; light violet: the corresponding probabilities for ordered crystalline $CBr_4$\cite{tlcbr4} ('x' scale multiplied by 1.1). Lower panels [(c) and (d)],  black solid lines: 1:2 [(c)] and 1:1 [(d)] arrangements; red solid lines with circles: the corresponding probabilities as calculated from the average structure; grey tone lines: 2:3 [(c)] and 3:3 [(d)] arrangements; blue solid lines with crosses: the corresponding probabilities as calculated from the average structure; olive line: the centre-centre prdf (shifted along the 'y' axis and re-scaled for a better visibility).
}
\end{center}
\end{figure}

\Fref{fig:rey} shows the probabilities of each class as a function of the distance between the centres of two molecules. A general observation is that only 2:2 and 1:3 arrangement play major roles in the system, others are marginal. It should be noted that there are distance ranges where the carbon-carbon prdf contains only very few atoms in a given distance bin (which is quite normal for a crystal)-- i.e., the definition of any 'distribution' would be meaningless. This is the reason why between 7.5 and 8.2 {\AA } no intensities are shown in \fref{fig:rey}.

Edge-edge arrangements are the most probable ones: we can find them in great numbers around the maxima of the carbon-carbon prdf. However, the first maximum of the 2:2 distribution appears at shorter distances than the position of the first maximum of the C-C prdf. Comparing with the monoclinic phase of $CBr_4$, the 2:2 arrangement there is much less preferred at shorter distances; instead, the phase change from the ordered to the plastic crystalline phase of $CBr_4$ is explained by the transformation of several classes into 2:2 in the plastic crystalline phase\cite{tlcbr4}. The large number of 2:2 pairs at the 'contact' (very short) distance seems to be a distinct feature of tetragonal $CI_4$. At medium range (above 8 {\AA }), the two crystalline materials follow similar characteristics.

Similarly, the Apollo-like 1:3 arrangements appear in the first shell with quite a large probability; in contrast to the 2:2 class, 1:3 pairs are most frequent around the exact position of the first maximum of the C-C prdf. According to the angular analysis above, the iodine close to the line connecting neighbouring carbon atoms would suggest a longer centre-centre distance (6.6 {\AA }) than it is observed (6.5 {\AA }): this kind of distortion is shown in \fref{fig:rey}. As a consequence of the crystalline symmetry, the 1:3 arrangement becomes very probable also at 13.3{\AA}, which corresponds to the distance of the body diagonal of the Bravais-cell. (The slight difference between observed expected distances seems to propagate along the body diagonal.) We found similar characteristics in the monoclinic phase of $CBr_4$ at medium range. A general observation concerning the two most important orientations: along the body diagonal, 1:3 arrangements dominate whereas along the 'a' vector, 2:2 pairs are more abundant.

Although their role is marginal, it is still worth spending a few words on the remaining four classes of orientations. 1:1 and 3:3, and 1:2 and 2:3 arrangements are in phase (not like in the corresponding liquid phases\cite{szilviccl4}), which can only be explained by a well defined crystalline structure. These pairs appear with relatively larger probabilities where the intensity of the carbon-carbon prdf is low. The only exception is the 2:3 arrangement that shows up with a considerable probability in the region of the closest molecule-molecule distance: their presence may be taken as deviation from the (nearly uniform) 2:2 alignment and thus, they may contribute to the observed diffuse scattering intensity. 

\section{Conclusions}

We performed total neutron scattering diffraction measurements on the tetragonal crystalline phase of carbon tetraiodide. The experimental total scattering structure factor was modeled successfully by the RMCPOW\cite{rmcpow} algorithm. As seen unambiguously from displaying atomic positions (see \Fref{fig:reduced}), the observed diffuse scattering comes from an enhanced positional disorder of iodine atoms, i.e. from molecular librations. The previously suggested\cite{pohl} intermolecular iodine-iodine correlation at 4.05 {\AA} has been found, although its manifestation in terms of the prdf is not very distinct. The carbon-iodine correlation centered around 4.5 {\AA} has been found to be significant; this feature of the corresponding prdf results from 'Apollo-like' (1:3) mutual arrangements. The cosine distribution of the C-I-C intermolecular angles suggests that iodine atom positions close to, but not on, the intermolecular carbon-carbon connecting line are preferred. Analyses of mutual orientations of two molecules showed that edge-edge and corner-face (Apollo-like) near neighbours are the most important and also, that these types of arrangements remain distinct at larger intermolecular distances, too. This behavior is completely different from that found for the monoclinic phase of $CBr_4$\cite{tlcbr4}.

\ack
LT is grateful to Anders Mellerg{\aa}rd and Per Zetterstr\"om for sharing their knowledge on RMCPOW. Financial support was provided by the Hungarian National Research Fund (OTKA), via grant No. 083529.  

\section*{References}
\bibliographystyle{unsrt}
\bibliography{ci4diffuse}

\end{document}